\documentclass[pra,twocolumn,superscriptaddress,a4paper,nofootinbib]{revtex4-2}
\usepackage[utf8]{inputenc}
\usepackage{graphicx}
\usepackage{enumitem}
\usepackage{amsmath}
\usepackage{amsfonts}
\usepackage[normalem]{ulem}
\usepackage[T1]{fontenc}

\usepackage[colorlinks,breaklinks,citecolor=blue,linkcolor=black,urlcolor=black]{hyperref}

\newcommand{\ket}[1]{|{#1}\rangle}

\newcommand{\Hop}{\hat{H}}

\begin{document}

\title{On the direct diagonalization method for a few particles trapped in harmonic potentials}

\author{Abel Rojo-Franc{\`a}s}
\affiliation{Departament de F{\'i}sica Qu{\`a}ntica i Astrof{\'i}sica, Facultat de F{\'i}sica, Universitat de Barcelona, E-08028 Barcelona, Spain}
\affiliation{Institut de Ci{\`e}ncies del Cosmos, Universitat de Barcelona, ICCUB, Mart{\'i} i Franqu{\`e}s 1, E-08028 Barcelona, Spain}
\author{Felipe Isaule}
\affiliation{School of Physics and Astronomy, University of Glasgow, Glasgow G12 8QQ, United Kingdom}
\author{Bruno Juli{\'a}-D{\'i}az}
\affiliation{Departament de F{\'i}sica Qu{\`a}ntica i Astrof{\'i}sica, Facultat de F{\'i}sica, Universitat de Barcelona, E-08028 Barcelona, Spain}
\affiliation{Institut de Ci{\`e}ncies del Cosmos, Universitat de Barcelona, ICCUB, Mart{\'i} i Franqu{\`e}s 1, E-08028 Barcelona, Spain}

\begin{abstract}
We describe a procedure to systematically improve 
direct diagonalization results for few-particle systems 
trapped in one-dimensional harmonic potentials interacting 
by contact interactions. We start from the two-body problem to 
define a renormalization method for the interparticle interactions. The procedure is 
benchmarked with state-of-the-art numerical results for 
three and four symmetric fermions.

\end{abstract}

\maketitle

\section{Introduction}

Ultracold atomic gases laboratories provide versatile 
setups for the quantum simulation of a large number of phenomena 
in condensed matter and many-body 
quantum physics~\cite{lewenstein_ultracold_2007,bloch_many-body_2008}. 
These setups allow to study the onset of many-body quantum physics, 
in experiments where the systems can be made to transit from the 
few-body regime~\cite{PhysRevLett.108.075303,doi:10.1126/science.1240516,zurn_pairing_2013} into 
the many-body one, e.g. 
Bose-Einstein condensates~\cite{RevModPhys.71.463}. 

Simultaneously, the theoretical and numerical efforts to understand 
the transition from the few- to the many-body problem have flourished 
in a number of well-consolidated techniques, such as Monte 
Carlo methods~\cite{10.1007/BFb0104529}, 
tensor networks~\cite{doi:10.1080/14789940801912366}, mean-field 
approaches~\cite{RevModPhys.71.463}, coupled-cluster method~
\cite{grining_crossover_2015}, direct diagonalization 
techniques~\cite{PhysRevA.75.013614, Ravent_s_2017} and, more recently, 
machine learning ones~\cite{doi:10.1126/science.aag2302}. All of them 
have their pros and cons, all of them bear inherent approximations 
which make them useful only in certain conditions, e.g. low dimensions, 
mild interaction regimes, few particles, etc. 

In this work, we concentrate on direct diagonalization 
techniques mostly used for particles trapped in a 1D harmonic 
potential, e.g.~\cite{sowinski_few_2013,rojo-francas_static_2020}. In this method 
the idea is simple, one needs to build the many-body Hamiltonian on
a suitable basis and diagonalize it "exactly". The method does not provide 
exact results due to the truncations made on 
the Hilbert space. The usual procedure runs as follows: 1) fix the 
number of particles, $N$, to be either bosons or fermions, or mixtures. 
Then, 2) truncate the single-particle basis to $M$ modes, and 3) build 
the corresponding many-body basis performing a truncation on the total 
energy of the non-interacting many-body states~\cite{sowinski_efficient_2019}. 

This technique has been recently used to study small 
1D bosonic mixtures~\cite{Garc_a_March_2014, sowinski_one-dimensional_2019}, 
fermionic systems~\cite{rojo-francas_static_2020}, 
2D bosonic systems with and without 
spin-orbit coupling~\cite{PhysRevA.96.043614,PhysRevA.101.043619}. 
Even though direct diagonalization calculations are limited to a small 
number of particles, they offer several advantages 
compared to other approaches. First, they provide access to a large portion 
of the energy spectrum. In particular, they give the full solution of these 
states, including the eigenstates, excitation properties and spectral 
functions. In contrast, many approaches are restricted to a few 
ground-state properties. In addition, depending on the size of the truncated 
Hilbert space, direct diagonalization can be easily used to perform 
time-dependent calculations.

An issue that remains elusive concerns the way to perform 
extrapolations on the number of single-particle modes, $M$. This issue 
has been tackled in previous works, notably in 
Refs.~\cite{ernst_simulating_2011,jeszenszki_accelerating_2018} 
and, specifically for harmonic traps, 
in Refs.~\cite{lindgren_fermionization_2014,dehkharghani_quantum_2015,grining_many_interacting_2015}. 
These studies propose a heuristic scheme to perform the extrapolation of 
the results computed for a finite $M$ to the $M \to \infty$ limit.
Other studies address this problem by describing an effective interaction, see Ref.~\cite{rotureau_interaction_2013,rammelmuller_a_modular_2022}.

In this work, we describe a procedure to systematically perform 
the $M \to \infty$ limit of few-particle properties, e.g. we 
consider eigenenergies and density profiles. The procedure is benchmarked
with state-of-the-art few-body calculations for $N=3$ and $N=4$ 
fermionic SU($N$) symmetric systems~\cite{laird_su_2017}. Our method shows 
an outstanding performance, providing results with less than $1\%$ 
of discrepancy with the exact ones for $N=3$ and $N=4$ particles with 
as few as 20 modes for interaction strengths $g$ in the whole $0\to\infty$ range.

Our work is organized as follows. In Sec.~\ref{sec:m} we 
describe the Hamiltonian. Then, in 
Sec.~\ref{sec:ap} we revise the analytical solution of 
the two-particle case~\cite{busch_two_1998}, which is then used in the 
extrapolation algorithm, described in this section. In 
Sec.~\ref{sec:r2p} we present how the 
procedure works for just two particles. In this case, the approach is 
exact and allows one to understand how to use it for more particles. 
In Sec.~\ref{sec:emp} we consider the few-particle scenario. There we compare with the exact results of Ref.~\cite{laird_su_2017} 
for the lower part of the energy spectrum obtained for three and four SU($N$) particles 
and we also report the energy predictions for five and six. We also discuss the 
correction on the single-particle density. Finally, in Sec.~\ref{sec:sc} we 
present a summary and the main conclusions of our work.

\section{Model}
\label{sec:m}

Let us consider a system composed of a few particles, bosons or fermions, 
with a number of internal states, trapped in a one-dimensional harmonic 
oscillator (HO) potential. We assume that the interaction is properly 
described by a contact potential, as is the case for many ultracold 
atomic gases experiments, see for instance 
the reviews~\cite{bloch_many-body_2008,lewenstein_ultracold_2007}. 
In first quantization the Hamiltonian of the system for $N$ particles is 
\begin{equation}
    \Hop=\sum_{i=1}^N\left[-\frac{\hbar^2}{2m}\frac{\partial^2}
    {\partial x_i^2}+\frac{m\omega^2}{2}x_i^2\right]
    +\sum_{\alpha,\beta}g_{\alpha\beta} 
    \sum^{N_\alpha,N_\beta}_{i < j}\delta(x_i-x_j)\,,
    \label{sec:model;eq:HO}
\end{equation}
where $g_{\alpha\beta}$ is the interaction strength between the particles 
in internal states $\alpha$ and $\beta$, and $N_\alpha$ is the number of 
particles in the internal state $\alpha$.

By choosing  the  HO eigenfunctions as the single-particle basis, the 
HO part of the Hamiltonian (\ref{sec:model;eq:HO}) is diagonal with 
eigenvalues $\epsilon_{n_i}=(n_i+1/2)\hbar \omega$. $n_i$ is the index of 
the HO wavefunction of the state $\ket{i}$. This state has a spatial and 
a spin component: $\ket{i}=\ket{\Phi_{n_i}\chi_{s_i}}$, where $\Phi_{n_i}$ 
is the $n_i$-th HO wavefunction and $\chi_{s_i}$ is the internal state 
wavefunction of internal state $s_i$. 

In the HO basis, the two-body matrix elements of the interacting 
part of the Hamiltonian (\ref{sec:model;eq:HO}) are 
expressed as~\cite{rojo-francas_static_2020},
\begin{equation}
    v_{ij,kl}=g_{s_is_j}\,\delta_{s_i,s_k}\delta_{s_j,s_l}\int dx\,\Phi_{n_i}(x)\Phi_{n_j}(x)\Phi_{n_k}(x)\Phi_{n_l}(x)\,,
     \label{sec:model;eq:vijkl}
\end{equation}
where $\Phi_n(x)$ are the eigenfunctions of the HO Hamiltonian for the 
energy level $n$, which are real in one dimension. Note that we have used the 
orthogonality of the spin functions: 
$\langle \chi_{s_i}\chi_{s_j}|\chi_{s_k}\chi_{s_l}\rangle=
\delta_{{s}_i,{s}_k}\delta_{{s}_j,{s}_l}$. We also stress
that the interaction does not affect the spin of the particles.
We numerically calculate the integral (\ref{sec:model;eq:vijkl}) using 
the procedure presented in Ref.~\cite{rojo-francas_static_2020}.

The full Hamiltonian (\ref{sec:model;eq:HO}) in second quantization reads
\begin{equation}
\Hop=\sum_i \epsilon_{n_i} \hat{a}^\dagger_i\hat{a}_i + \frac{1}{2}\sum_{ijkl}v_{ij,kl}\hat{a}^\dagger _i \hat{a}^\dagger_j \hat{a}_l \hat{a}_k\,.
     \label{sec:model;eq:HOsq}
\end{equation}
where $\hat{a}_i^\dagger$ ($\hat{a}_i$) creates (annihilates) a particle in 
the single-particle state $\ket{i}$.
\section{Correction of the truncated results}
\label{sec:ap}

As explained above, using direct diagonalization techniques one numerically obtains 
the lower-energy eigenvalues and corresponding eigenstates in a truncated Hilbert 
space. Importantly, these techniques fall within the variational method, 
i.e. they do produce in all cases upper bounds to the corresponding exact 
eigenvalues. In practice, one has to truncate the single particle basis to 
a finite number of modes $M$ (for details see~\cite{rojo-francas_static_2020}).
In this work, we additionally truncate the many-body basis up to a 
non-interacting energy $E_\mathrm{max.}(M)$. We discuss this in detail in Sec.~\ref{sec:ap;subsec:trunc}.
 
Being variational, increasing the value of $M$, thus enlarging the Hilbert 
space lowers the value of the upper bound. However,
these approximate results can deviate
considerably 
from the exact values, especially for strong interactions.
Building upon the ideas proposed 
in Refs.~\cite{ernst_simulating_2011,jeszenszki_accelerating_2018}, in the 
following we detail a procedure to improve the truncated results by 
correcting the potential using the known two-body solutions. 

\subsection{Two-particle exact solution}
\label{sec:ap;sub:2B}

We start examining the problem of two particles in a HO interacting 
with a contact potential of strength $g$, which can be solved analytically. 
We follow the derivation in Ref.~\cite{busch_two_1998} but 
restricted to one dimension. 

For two particles the Hamiltonian (\ref{sec:model;eq:HO}) in HO units, reads
\begin{equation}
    \Hop=-\frac{1}{2}\frac{\partial^2}{\partial x_1^2}-\frac{1}{2}\frac{\partial^2}{\partial x_2^2}+\frac{x_1^2}{2}+\frac{x_2^2}{2}+g\,\delta(x_1-x_2)\,.
\end{equation}
Working with the center-of-mass (c.m.) and relative coordinates 
$X_{\mathrm{c.m.}}=(x_1+x_2)/\sqrt{2}$ and 
$x=(x_1-x_2)/\sqrt{2}$, respectively, the 
Hamiltonian can be written as $\Hop=\Hop_{\mathrm{c.m.}}+\Hop_{\mathrm{rel}}$.
The center-of-mass Hamiltonian is simply a HO with 
eigenvalues $E_{\mathrm{c.m.}}=(n_\mathrm{c.m.}+1/2)\hbar \omega$. On the 
other hand, the Schrödinger equation for the relative part reads
\begin{equation}\label{Shrod.eq}
\left(\Hop_{\mathrm{HO}}+\frac{g}{\sqrt{2}}\delta(x)\right)\Psi(x)=
E_\mathrm{r}\Psi(x)\,.
\end{equation}
$\Hop_{\mathrm{HO}}$ is the HO Hamiltonian for the relative coordinate. 
Expanding the relative wavefunction in the HO basis
\begin{equation}
    \Psi(x)=\sum_m c_m\Phi_m(x)\,,
\end{equation}
and projecting the state on $\Phi_n(x)$, we obtain
\begin{equation}
    c_n=A\frac{\Phi_n(0)}{E_n-E_\mathrm{r}}\,,
\end{equation}
where $A=-\frac{g}{\sqrt{2}}\sum c_m \Phi_m(0)$ is a constant that does 
not depend on $n$. From these we get 
\begin{align}
    &1+\frac{g}{\sqrt{2}}\sum_{n=0}^\infty \frac{\Phi_n(0)\Phi_n(0)}{E_n-E_\mathrm{r}}=0\,.
\end{align}
Using the explicit values of the wavefunctions $\Phi_n$ at the 
center of the trap and noting that only the even $n$ terms 
contribute to the sum, we obtain
\begin{equation}
\label{sec:ap;sub:2B;infinite_sum}
    -\frac{1}{g}\sqrt{\frac{\omega\hbar^3}{m}}=\sum_{n'=0}^\infty f(n',\nu)\,,
\end{equation}
where we have performed the change $n= 2n'$ and
\begin{equation}
    f(n',\nu)=\frac{1}{2\sqrt{2\pi}}\frac{(2n')!}{4^{n'}(n'!)^2(n'-\nu)}\,.
\end{equation}
For convenience, here we have defined $E_\mathrm{r}= (2\nu +1/2)\hbar\omega$, 
and thus, $\nu = E_\mathrm{r}/(2\hbar\omega)-1/4$. From here on, the 
interaction strengths will be expressed in harmonic oscillator units, 
i.e. $\sqrt{\omega\hbar^3/m}$. The sum in 
(\ref{sec:ap;sub:2B;infinite_sum}) can be solved in closed-form, 
resulting in~\cite{busch_two_1998}
\begin{equation}
   \frac{\Gamma(-\nu)}{\Gamma(1/2-\nu)}=-\frac{2^{3/2}}{g}\,.
   \label{sec:ap;sub:2B;eq:exact_g}
\end{equation}
Eq.~(\ref{sec:ap;sub:2B;eq:exact_g}) 
determines the energies of the relative system which, in combination 
with the c.m. energies $E_{\mathrm{c.m.}}$, define the full energy spectrum 
of the two-body problem as a function of $g$.

\subsection{Truncation of the exact two-body solution}
\label{sec:ap;sub:2Btrunc}

To connect the exact two-body solution with the upcoming 
truncated basis for more particles, we truncate the 
sum~(\ref{sec:ap;sub:2B;infinite_sum}) to the subspace with the 
first $M$ modes. Due to the change $n=2n'$, we define 
$\mathcal{M}=\lfloor(M-1)/2\rfloor$, where $\lfloor x\rfloor$ is the 
floor function of $x$. 
In this case, the sum takes the form
\begin{align}
\label{sec:ap;sub:2B;finite_sum}
    -\frac{1}{g'}\equiv &\sum_{n'=0}^\mathcal{M}  f(n',\nu)\\
    =&\frac{\Gamma (-\nu)}{2^{3/2}\Gamma \left(\frac{1}{2}-\nu\right)}-\Gamma \left(\mathcal{M}+\frac{3}{2}\right) \Gamma (\mathcal{M}-\nu+1)\nonumber\\ 
    \times&\frac{ \, _3\tilde{F}_2\left(1,\mathcal{M}+\frac{3}{2},\mathcal{M}-\nu+1;\mathcal{M}+2,\mathcal{M}-\nu+2;1\right)}{2\pi\sqrt{2 }}\,, \nonumber
\end{align}
where $_3 \tilde{F}_2$ is a hypergeometric regularized function. 
In the limit of the full basis $\mathcal{M}\rightarrow\infty$, 
Eq.~(\ref{sec:ap;sub:2B;finite_sum}) recovers Eq.~(\ref{sec:ap;sub:2B;eq:exact_g}).

Note that we have introduced a truncated interaction strength $g'$, 
which depends on the size of the truncated basis $\mathcal{M}$. Indeed, 
for a specific two-body energy of the relative system $E_r$, $g'$ 
equals the physical interaction strength $g$ only in the limit
$\mathcal{M}\rightarrow\infty$.

Our main objective is to find a relation between the 
exact $E_r$ given by Eq.~(\ref{sec:ap;sub:2B;eq:exact_g}),
and the truncated solution.
To this end, for a fixed value of the relative system energy $\nu$ we 
separate the infinite sum in Eq.~(\ref{sec:ap;sub:2B;infinite_sum}) into 
two terms
\begin{equation}
    -\frac{1}{g(\nu)}= \sum_{n'=0}^\mathcal{M}f(n',\nu) + \sum_{n'=\mathcal{M}+1}^\infty f(n',\nu)\,,
\end{equation}
where the first term in the right-hand-side corresponds to the sum 
in Eq.~(\ref{sec:ap;sub:2B;finite_sum}), and thus, it can be written 
as $1/g'$. Analogously, by defining an interaction strength correction 
$g_c$ as $\sum_{n'=\mathcal{M}+1}^\infty f(n',\nu)=1/g_c$, we can write
\begin{equation}
    \frac{1}{g(\nu)}=\frac{1}{g'(\mathcal{M},\nu)}-\frac{1}{g_c(\mathcal{M},\nu)}\,,
    \label{g_renormalization}
\end{equation}
which connects the physical interaction strength $g$ with its truncated 
counterpart $g'$ for a chosen energy of the relative system. Note that 
this equation has a similar form to those used to regularize two-body 
interactions in quantum gases~\cite{stoof_ultracold_2009}.

The value of $g_c$ for a chosen number of modes $M$ can be obtained from Eqs.~\eqref{sec:ap;sub:2B;eq:exact_g} and~\eqref{sec:ap;sub:2B;finite_sum}. We find
\begin{align}\label{1/g_cgeneral}
    \frac{1}{g_c}=&\Gamma \left(\mathcal{M}+\frac{3}{2}\right) \Gamma (\mathcal{M}-\nu+1) 
    \\\times&\frac{\, _3\tilde{F}_2\left(1,\mathcal{M}+\frac{3}{2},\mathcal{M}-\nu+1;\mathcal{M}+2,\mathcal{M}-\nu+2;1\right)}{2\pi\sqrt{2 }}\,, \nonumber
\end{align}
which depends on both the energy and the number of modes. However, 
because the terms in $\mathcal{M}$ become dominant for large $\mathcal{M}$, 
the dependence of $1/g_c$ on the 
number of modes is much more relevant than that on the energy.

Eq.~(\ref{1/g_cgeneral}) enables us to connect the truncated results with 
the exact solution. However, the numerical evaluation of Eq.~\eqref{1/g_cgeneral} 
can be very time consuming due to the hypergeometric functions. To speed 
up the numerical calculations we propose an approximation for 
Eq.~\eqref{1/g_cgeneral}. First, by using Stirling's asymptotic 
formula we have that
\begin{equation}
    \begin{split}
    \frac{1}{g_c (\mathcal{M},\nu)} & =\frac{1}{2\sqrt{2\pi}} \sum_{n'=\mathcal{M}+1}^\infty\frac{(2n')!}{2^{2n'}(n'!)^2(n'-\nu)}\\
    &\simeq \frac{1}{2\sqrt{2\pi}} \sum_{n=\mathcal{M}+1}^\infty  \frac{1}{\sqrt{n'\pi}(n'-\nu)}\,.
    \end{split}
\end{equation}
Then, we turn this summation into an integral by using the 
Euler-McLaurin formula. We obtain
\begin{equation}\label{Euler-McLaurin}
\begin{split}
    \frac{2\pi \sqrt{2}}{g_c}\simeq &\frac{1}{\sqrt{\nu}}\ln\left(\frac{\sqrt{\mathcal{M}+1}+\sqrt{\nu}}{\sqrt{\mathcal{M}+1}-\sqrt{\nu}}\right)\\
    &+\frac{1}{2\sqrt{\mathcal{M}+1}(\mathcal{M}+1-\nu)}\\
    &\times\left(1+\frac{1}{12(\mathcal{M}+1)}+\frac{1}{6(\mathcal{M}+1-\nu)}\right) \,.
\end{split}
\end{equation}
This approximation for $g_c$ has an error of less than $1\%$ with 
respect to its exact value for $\nu < \mathcal{M}$ and can be up 
to $\sim 10^6$ times faster to evaluate numerically. The results 
shown in the rest of this work, Secs.~\ref{sec:r2p} and \ref{sec:emp}, 
are obtained using this approximation.

In principle, Eqs.~\eqref{1/g_cgeneral} and \eqref{Euler-McLaurin} can 
be evaluated for any value of $\mathcal{M}$ and $\nu$. However, both 
expressions have a pole at $\nu=\mathcal{M}+1$. Moreover,  for larger 
values of $\nu$, Eq.~\eqref{1/g_cgeneral} oscillates from $-\infty$ 
to $\infty$, while Eq.~\eqref{Euler-McLaurin} gives imaginary numbers. 
For this reason, these expressions only have useful values when 
$\nu\leq\mathcal{M}+1$.
This relates to the excitation energy and the number of modes 
used in the basis as $\Delta E/ \hbar\omega \leq M+1$, where 
$\Delta E$ is the difference between the energy $E$ and the 
energy of the non-interacting ground state. This indicates 
that to correct a state with a certain energy $E$, we must 
include in our basis at least all states with non-interacting 
energy equal or lower than $E$.

Eqs.~(\ref{g_renormalization}) and (\ref{1/g_cgeneral}) 
(or (\ref{Euler-McLaurin})) enable us to improve two-body 
calculations in a truncated space by correcting the truncated 
strength $g'$ to its physical value $g$. We employ this 
idea to correct calculations for more particles in the following.

\begin{figure}[t!]
    \centering
    \includegraphics[width=\columnwidth]{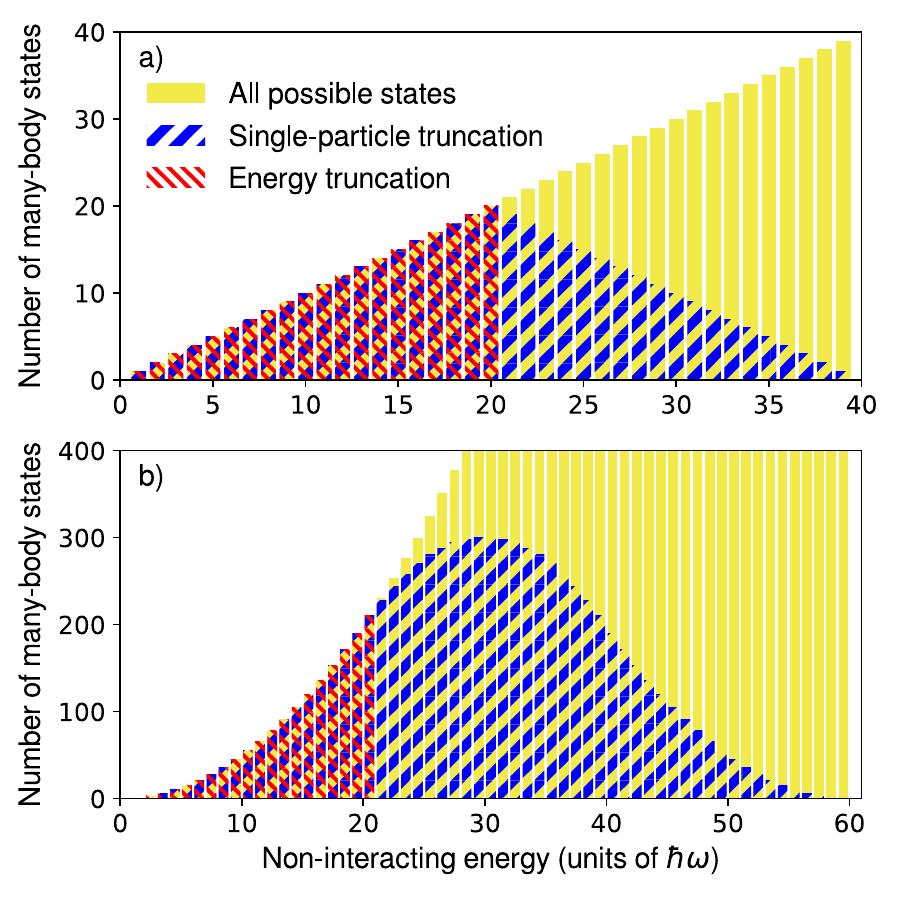}
    \caption{(Color online) Number of basis states for each value of the non-interacting energy. We consider 20 single-particle modes for $N=2$ (upper panel) and $N=3$ (lower panel). The yellow solid region is the number of states with less energy than $E=40~\hbar\omega$ and $E=60~\hbar\omega$ for two and three particles, respectively. The blue thick hatched region is the number of states that can be created with 20 single-particle states. The red thin hatched region is the number of states created with 20 single-particle states and with an energy truncation at $E=E_{\mathrm{max}}$.}
    \label{fig:states}
\end{figure}

\subsection{Truncation of the many-body basis}\label{sec:ap;subsec:trunc}

As mentioned, in systems with more than two particles one first needs to truncate the many-body basis to a finite number of HO states. To do this truncation, we choose a number of modes $M$ and then simply truncate the basis up to all the states with non-interacting energy smaller or equal than $E_\text{max}(M)$. This energy truncation enables us to greatly reduce the size of the many-body basis while maintaining the quality of the results~\cite{plodzien_numerically_2018,sowinski_efficient_2019}.

In systems composed of bosons or distinguishable particles, as the ones considered in this work, the optimal value for this maximum energy is~\cite{plodzien_numerically_2018}
\begin{equation}
    E_{\mathrm{max}}(M)/\hbar\omega=M-1+N/2\,,
    \label{sec:ap;subsec:trunc;eq:Emax}
\end{equation}
where $N$ is the number of particles\footnote{In fermionic systems the optimal maximum energy is
$E_{\mathrm{max}}/\hbar\omega(M)=M+E_F/\hbar\omega-\mathrm{max}\left(N_\alpha\right)$~\cite{plodzien_numerically_2018},
where $E_F$ is the Fermi energy and $N_\alpha$ is the number of particles in the 
internal state $\alpha$.}. Therefore, the basis is constructed as usual by discarding all the states with non-interacting energy larger than $E_{\mathrm{max}}$.

To illustrate the size of the truncated many-body basis, in Fig.~\ref{fig:states} we show the number of many-body states as a function
of the non-interacting energy. We compare the basis created by the energy truncation (red thin hatched region), explained before, with a standard truncation in the number of modes without the energy restriction (blue thick hatched region).
With the energy restriction one considers much fewer states than with a standard truncation. However, the energy truncation provides a complete basis up to $E_{\mathrm{max}}$. In contrast, a simple truncation in the number of modes results in an inconsistent basis where some non-interacting energy states are not considered (see difference between yellow solid and blue thick hatched regions). We provide additional details in Appendix~\ref{ap:ecb}.

Once we have created the truncated many-body basis, we numerically diagonalize the Hamiltonian for the lower part of the energy-spectrum. This diagonalization provides an approximate solution, in analogy to the truncated result for two particles [Eq.~(\ref{sec:ap;sub:2B;finite_sum})].
Afterward,  we correct these calculations by connecting the two-body sector of the truncated many-body results with the truncation in Sec.~\ref{sec:ap;sub:2Btrunc} for 
two particles [Eq.~(\ref{g_renormalization})]. Therefore, for each obtained eigenenergy, we can correct the truncated interaction strengths $g'$ to their physical values $g$ using Eq.~\eqref{1/g_cgeneral}.

We are able to perform this correction thanks to the energy truncation of the many-body basis. Indeed, because our basis includes all the center of mass and relative coordinate modes for energies up to $E_\text{max}$, the many-body basis contains all the modes considered in the exact two-body solution [see Sec.~\ref{sec:ap;sub:2Btrunc}]. In contrast, a standard truncation without the energy restriction does not fulfill this condition and thus it is not suitable for the correction procedure.

We stress that for the rest of the main text, all the results are obtained from truncations with the energy restriction (\ref{sec:ap;subsec:trunc;eq:Emax}).

\subsection{A practical procedure for the correction}
\label{sec:ap;sub:proc}

In practice, the algorithm to improve the results is sketched as,

\begin{enumerate}
\item Create the many-body basis of $N$ particles with $M$ 
harmonic oscillator modes and keep only the many-body states 
with a non-interacting energy smaller or equal than 
$E_{\mathrm{max}}$. This allows us to correct states with 
energy below $E_{\mathrm{max}}$.
\item Compute the Hamiltonan matrix for a chosen value of the interaction strength $g'_{\alpha\beta}$.
\item Diagonalize the Hamiltonian and obtain the eigenvalues.
\item For each eigenvalue $E$, use $\nu=(E/\hbar\omega-N/2)/2$ to compute the correction $1/g_c$ using Eq.~\eqref{1/g_cgeneral} or Eq.~\eqref{Euler-McLaurin}.
\item Assign the interaction strength associated to this eigenvalue $E$ using $g_{\alpha\beta}=g_{\alpha\beta}'/(1-g_{\alpha\beta}'/g_c)$.
\end{enumerate}

This procedure is exact for correcting the energy of two 
particles as we show in the following section. Interestingly, 
as we show in Sec.~\ref{sec:emp}, this method can 
successfully be used for more particles.

\section{Results for two particles}
\label{sec:r2p}

\begin{figure}[t]
    \centering
    \includegraphics[width=\columnwidth]{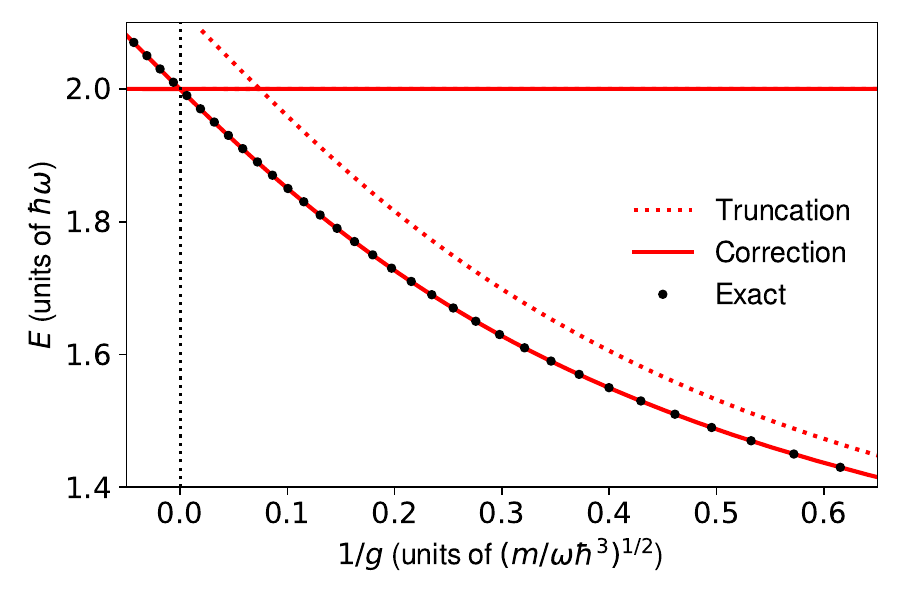}
    \caption{(Color online) Ground-state energy of the two particles system with SU(2) symmetry. The dotted line corresponds to the truncated results obtained with direct diagonalization using a basis of 20 single-particle modes, whereas the solid line corresponds to the corrected results using Eq.~\eqref{Euler-McLaurin}. The black circles correspond to the exact values of Eq.~\eqref{sec:ap;sub:2B;eq:exact_g}. The vertical dotted line indicates $1/g=0$. The horizontal line shows the Tonks energy for two particles.
    }
    \label{fig:2p_20}
\end{figure}

To illustrate how the correction procedure works, we first 
examine its application to the two-body problem.
In Fig.~\ref{fig:2p_20} we show the ground-state energy for 
two particles as a function of the interaction strength. We 
show results obtained with direct diagonalization, both with 
and without our correction scheme, and we compare them with 
the exact analytic results~\eqref{sec:ap;sub:2B;eq:exact_g}. 
We employ a small number of HO modes to better illustrate the 
improvement of the calculations. From now on, the values obtained
with the direct diagonalization without the correction will be
referred to as the truncated results and those with the correction
as the corrected ones.

We find that the correction (solid line) gives perfect 
agreement all digits with the exact results. In particular, reproducing 
the Tonks limit for two particles $E_\infty=2\hbar\omega$ 
for $1/g\to 0$. In contrast, the truncated calculation 
(dotted line) shows a noticeable deviation from the exact 
solution.
We have also checked that this agreement holds for the 
excited states. This can be expected, as the correction 
is exact for two particles [see Sec.~\ref{sec:ap;sub:2Btrunc}].

One interesting feature of our procedure is that for a 
strong truncated repulsion $g'$, the corrected physical 
strength $g$ becomes negative and corresponds to a strong attractive 
interaction. In Fig.~\ref{fig:2p_20}, this can be appreciated 
when the corrected energies cross from positive to negative 
$g$. Indeed, when we perform a truncated diagonalization 
calculation for $g'\to+\infty$, the resulting energy is 
greater than the Tonks solution for infinite 
repulsion~\cite{rojo-francas_static_2020}. Therefore, 
correcting the interaction strength, we obtain an 
attractive physical strength $g$ for an excited energy 
state in the attractive branch~\cite{ernst_simulating_2011}. 

As a consequence, we can map all the repulsive interacting 
regime $g>0$ with a finite range of $g'$. At the same time, the 
attractive regime $g<0$ cannot be mapped completely with a finite 
range of $g'$. With this correction, we can compute the correction 
for the weakly-interacting regime using $g'<0$. On the other hand, we 
can compute the correction for the strong attractive limit using $g'\gg0$

\begin{figure}[t]
    \centering
    \includegraphics[width=\columnwidth]{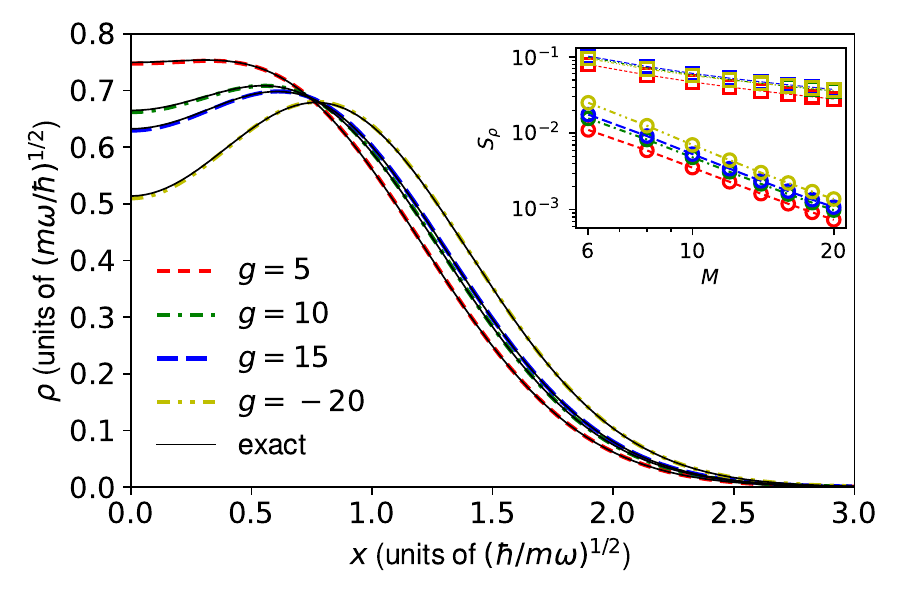}
    \caption{(Color online) Density profiles of two particles for different choices of interaction strengths $g$ (given in the labels in units of $(\omega\hbar^3/m)^{1/2}$).
    The thick lines correspond to profiles obtained with corrected direct diagonalization calculations with 20 single-particle modes, whereas the black thin lines are exact results given by Eq.~(\ref{eq:exact_wavefunction_2p}).
    The inset shows a correlation parameter $S_\rho$ between the exact profiles~(\ref{eq:exact_wavefunction_2p}) and the profiles obtained with the truncated diagonalizations (squares with thin lines) and the corrected ones (circles with thick lines) as a function of the number of single-particle modes $M$. 
    \label{fig:2p_density}}
\end{figure}

Our procedure can also be used to correct the interaction strengths 
associated with other properties, despite being exact only for 
correcting the energies of two particles. To illustrate this, in 
Fig.~\ref{fig:2p_density} we show the two-particle density 
profiles of the ground state for several values of $g > 0$. In addition, 
we also depict the density profile of the first relative excitation for 
a large attractive interaction.
We stress that the values of $g$ in the figure are those of the 
corrected interaction strengths. Therefore, the original truncated 
calculations were performed for truncated strengths $g'$ given by Eq.~\eqref{g_renormalization}. 

We compare our results with the exact profiles obtained
integrating the exact wavefunction~\cite{busch_two_1998} 
\begin{equation}\label{eq:exact_wavefunction_2p}
    \Psi (x_1,x_2)=A e^{-(x_1^2+x_2^2)/2}U\left(-\nu,\frac{1}{2},\frac{1}{2} (x_1-x_2)^2\right)\,,
\end{equation}
where $A$ is a normalization constant and $U(a,b,z)$ is the Tricomi function. All the parameters are
in harmonic oscillator units\\ \\

Our numerical calculations are in perfect agreement with the exact 
results, showing that our procedure also corrects the density profiles. 
In particular, the profile for the attractive strength 
$g/(\omega\hbar^3/m)^{1/2}=-20$ was obtained from a truncated 
repulsive $g'$, showing that the previously discussed change from a 
repulsive to an attractive interaction is indeed correct.

To quantify the accuracy of the correction we define a correlation 
parameter between two density profiles as
\begin{equation}
    S_\rho(\rho_1,\rho_2)=\frac{\int_{-\infty}^{\infty} |\rho_1(x)-\rho_2(x) |dx}{2N}\,.
\end{equation}
This parameter is zero when both profiles are equal and is one 
when both densities do not have any common region. In the inset of 
Fig.~\ref{fig:2p_density} we show the correlation between the exact 
density profiles and the ones obtained from direct diagonalization 
as a function of the number of modes. We show the value of the parameter $S_\rho$ with 
both the original truncated calculations using $g'=g$ 
(squares with thin lines) and with the corrected results (circles with thick lines). As 
expected, $S_\rho$ decreases with the number of modes for both methods, 
i.e. we are obtaining more precise results. In addition, not only $S_\rho$ has smaller
values for the corrected results, it also converges to zero faster than with the truncated ones.

\section{Extrapolation to many particles}
\label{sec:emp}

We now test our approach with more than two particles. We again stress 
that, in contrast to the two-particle case, our procedure is not exact 
for correcting the energy of more particles. And as we show in the 
following, our procedure greatly improves the truncated results.

\begin{figure}[t]
    \centering
    \includegraphics[width=0.95\columnwidth]{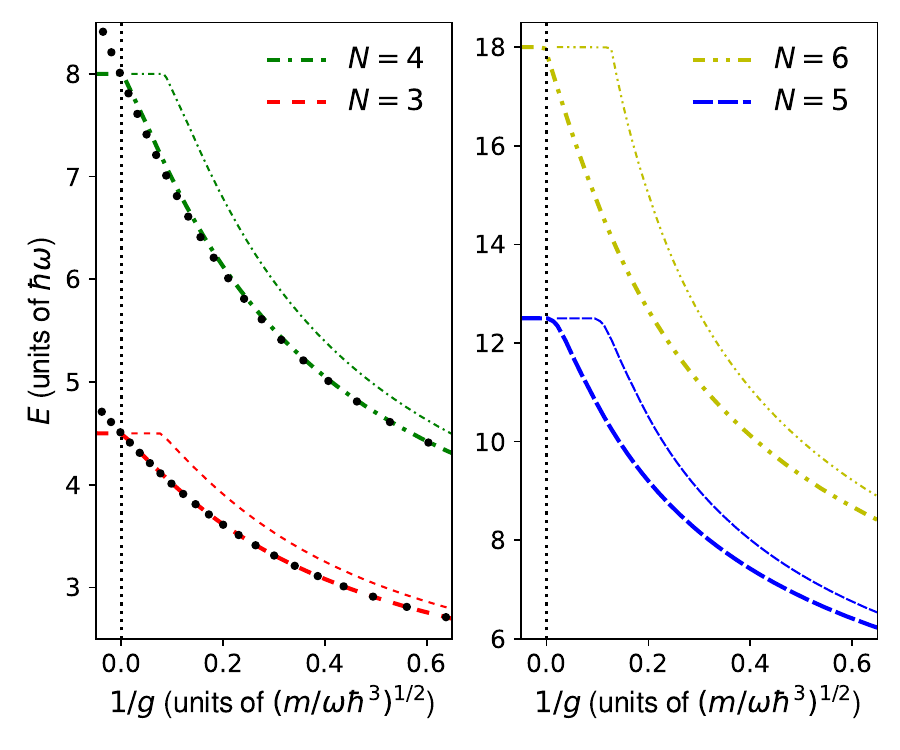}
    \caption{(Color online) Ground-state energies for $N=3$ to $N=6$ distinguishable particles computed using 20 single-particle modes. The thin lines correspond to the truncated calculations, whereas the thick lines correspond to the corrected energies obtained using Eq.~\eqref{Euler-McLaurin}. The black circles correspond to the exact values for $N=3$ and $N=4$~\cite{laird_su_2017}. The horizontal lines correspond to states with the Tonks energy for $N$ particles.}
    \label{fig:ground_state_many_particles}
\end{figure}

In Fig.~\ref{fig:ground_state_many_particles} we show the ground-state 
energy for $N=3$ to $N=6$ distinguishable particles with symmetric 
interactions $g=g_{\alpha\beta}$. We compare our results for three and 
four particles with exact solutions from Ref.~\cite{laird_su_2017}. As 
with the two-particle system, the original truncated calculations 
(thin lines) for $N=3$ and $N=4$ (left panel) show an important 
deviation from the exact results. In contrast, our corrected calculations 
(thick lines) show an almost perfect agreement with the exact solutions. 
We expect that this improvement holds for five and six particles (right panel).

The corrected calculations for $N \leq 5$ converge to the 
Tonks limit $E_\infty=N^2\hbar\omega/2$ for $1/g \sim 10^{-3}$,
whereas for $N=6$ the corrected energy reaches the Tonks limit at
$1/g \sim 10^{-2}$
. This larger deviation for six particles is due to the use of a small number of modes. Indeed,
for $N=6$ the Tonks energy $E_{\infty}=18\hbar\omega$ is too close to the limiting
energy for 20 modes. Nevertheless, this discrepancy is almost not appreciable 
in the figure. In contrast, the truncated calculations show a noticeable deviation,
saturating to the Tonks limit at a finite interaction strength in all cases.
\begin{figure}[t]
    \centering
    \includegraphics[width=0.95\columnwidth]{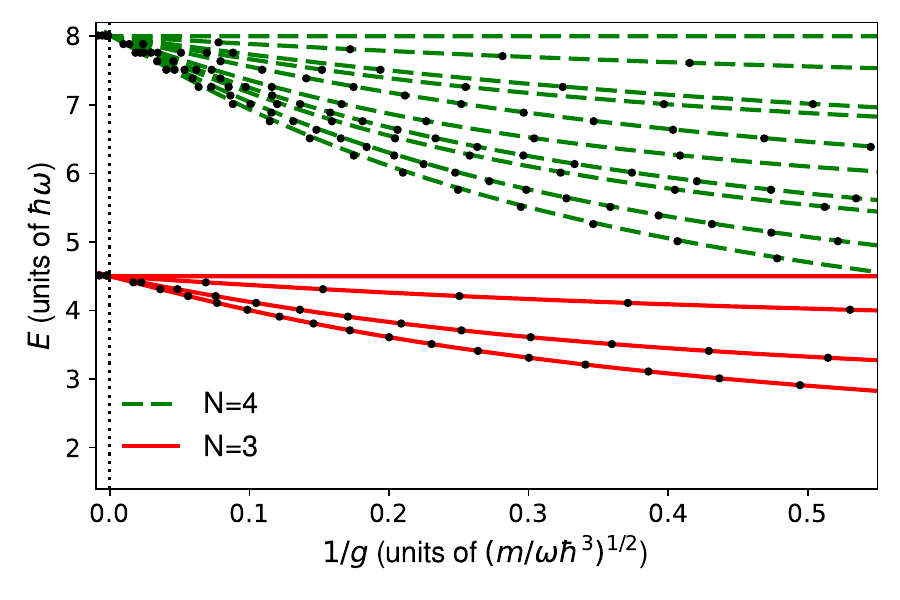}
    \caption{(Color online) Low-energy spectra for three and four particles, computed with 90 and 45 h.o. modes, respectively. The lines correspond to the corrected computations and the black circles to the exact results of Ref.~\cite{laird_su_2017}.}
    \label{fig:3p_4p_spectrum}
\end{figure}
In Fig.~\ref{fig:3p_4p_spectrum} we show the low-energy spectra of three 
and four distinguishable particles with symmetric interactions. We compare
our corrected calculations (lines) with the exact solutions (circles) from
Ref.~\cite{laird_su_2017}. 
We show the states that degenerate with the ground-state 
at the infinite interaction limit. Our correction has a great 
accuracy for three and four particles. The discrepancies are 
slightly larger for four particles. However, these discrepancies are
difficult to see in the figure. We provide an additional discussion on the dependence of the energy on the number of modes in Appendix~\ref{app:con}.\\
\begin{figure}[t]
    \centering
    \includegraphics[width=0.95\columnwidth]{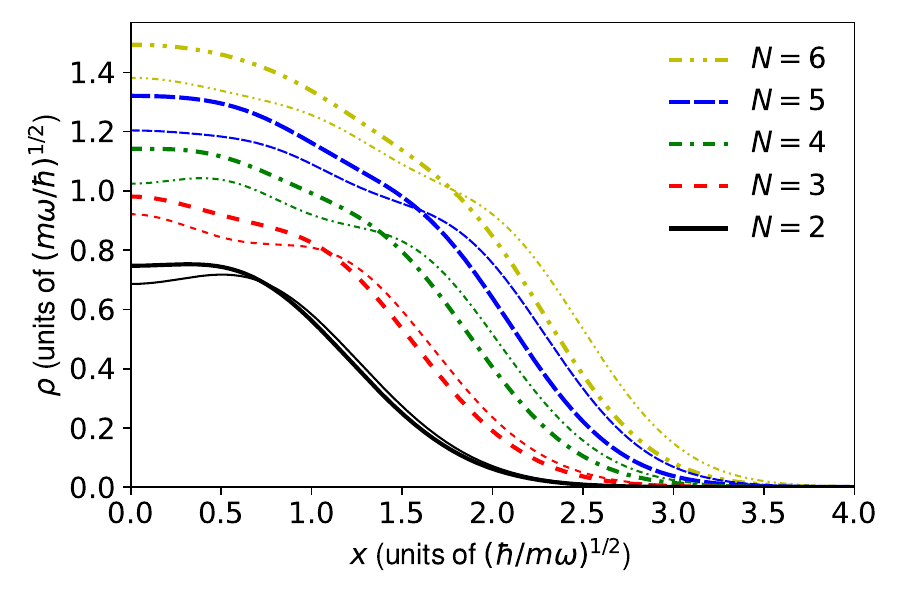}
    \caption{(Color online) Ground state density for several particles, where the thin lines are 
    the results of the truncated results at $g'/(\omega\hbar^3/m)^{1/2}=5$ and the thick lines are the corrected results at $g/(\omega\hbar^3/m)^{1/2}=5$.}
    \label{fig:density_many_particles}
\end{figure}
Finally, in Fig.~\ref{fig:density_many_particles} we show the density of the ground state for several
particles for a repulsive interaction strength $g/(\omega\hbar^3/m)^{1/2}=5$. We also show 
the density profiles obtained with the truncated calculation with $g'/(\omega\hbar^3/m)^{1/2}=5$ 
in order to compare the effect of the correction. For any number of particles, the corrected
density has a larger value at the center of the trap, while the tail has a smaller one. 
The densities corresponding to the truncated results are closer to the
Tonks density profile than the corrected ones, i.e. the 
truncated profile has the peaks corresponding to the density profile
of the infinite interacting limit whereas the corrected ones do not have it.
The differences between the truncated and corrected densities increase with the number
of particles. These differences can be quantified using the correlation parameter $S_\rho (\rho_t,\rho_c)$ 
between the truncated density $\rho_t$ and the corrected density $\rho_c$. 
This parameter increases (in general) as the number of particles increases, i.e.
$S_\rho (\rho_t,\rho_c)= 0.027$, $0.041$, $0.052$, $0.053$
and $0.050$ for two, three, four, five, and six particles, respectively.

\section{Summary and conclusions}
\label{sec:sc}

We have presented a well-defined procedure to extrapolate truncated 
direct diagonalization calculations for a few particles trapped in 
one-dimensional harmonic potentials. By employing the known two-body
solution, we can correct calculations truncated to a finite number of 
modes $M$ to the limit of the full basis $M\to\infty$. In contrast to previous 
literature, \cite{ernst_simulating_2011}, our method is not heuristic and 
does not require matching of the computed energies to the Tonks-Girardeau 
limit. In our case, we extrapolate the results by renormalizing 
the value of the interaction strength using only two-body information. 

We have found that this extrapolation procedure enables us to compute 
the low-energy spectrum of three and four distinguishable particles with an 
error of less than 1\% compared to exact solutions, even using a small number 
of single-particle modes. Furthermore, calculations for five and six 
particles correctly saturate to the Tonks limit. This suggests that, at 
least, by using the extrapolation we can provide a good qualitative 
description of systems with more than four particles.

The presented extrapolation is not constrained to certain particle 
statistics, interactions, or the number of particles. Therefore, this method 
can be applied to a plethora of scenarios, such as mixtures of bosonic 
and fermionic atoms, asymmetric interaction strengths, among others. 
This makes this method a good tool to study impurity physics and systems 
with broken SU($N$)-symmetry. In addition, the extrapolation could make 
accurate direct diagonalization studies with up to eight or ten particles 
accessible, bridging the gap between few- and many-body physics.

\begin{acknowledgements}
We thank Prof. Joan Martorell for his support in 
all aspects reported in this work. We also thank Emma Laird for sending
us her results for SU($N$) fermions from Ref.~\cite{laird_su_2017}. 
This work has been funded by Grant No. PID2020-114626GB-I00 from the MICIN/AEI/10.13039/501100011033. F.I. acknowledges funding from EPSRC 
(UK) through Grant No. EP/V048449/1. We acknowledge financial support 
from Secretaria d’Universitats i Recerca del Departament d’Empresa i 
Coneixement de la Generalitat de Catalunya, co-funded by the European Union
Regional Development Fund within the ERDF Operational Program of 
Catalunya (project QuantumCat, ref.
001-P-001644)
\end{acknowledgements}

\appendix

\section{Construction of the many-body basis}\label{ap:ecb}
\begin{table}[t]
\begin{ruledtabular}
    \centering
    \begin{tabular}{lrr}
        \multicolumn{1}{c}{\textrm{Particles and}} & \multicolumn{1}{c}{\textrm{Standard}} & \multicolumn{1}{c}{\textrm{Energy}} \\ 
        \multicolumn{1}{c}{\textrm{modes}} & \multicolumn{1}{c}{\textrm{truncation}} & \multicolumn{1}{c}{\textrm{truncation}} \\ 
        \colrule
         $N$=2, $M$=20 & 400 & 210 \\ 
        $N$=3, $M$=20 &8 000 & 1 540 \\ 
        $N$=3, $M$=90 &729 000 & 125 580 \\ 
        $N$=4, $M$=20 & 160 000& 8 855 \\ 
        $N$=4, $M$=45 & 4 100 625& 194 580 \\ 
        $N$=5, $M$=20 &3 200 000 & 42 504 \\ 
        $N$=6, $M$=20 & 64 000 000& 177 100 \\ 
    \end{tabular}
    \caption{
    Number of states in the many-body basis for different numbers of particles $N$ and harmonic oscillator modes $M$. We show the dimension of the Hilbert space without the energy restriction (standard truncation) and with the energy restriction (energy truncation) using $E_\mathrm{max}$ of Eq.~(\ref{sec:ap;subsec:trunc;eq:Emax}).}
    \label{tab:Hilbert_dimension}
\end{ruledtabular}
\end{table}
To construct the many-body basis we employ single-particle states $n$ of the HO Hamiltonian. For $N$ distinguishable particles, as considered in this work, we can write one state in such basis as
\begin{equation}
    \ket{\Psi}=\ket{n_1,n_2,...,n_N}\,,
\end{equation}
where $n_i$ is the HO index of particle $i$.
\begin{figure}[t]
    \centering
    \includegraphics[width=0.95\columnwidth]{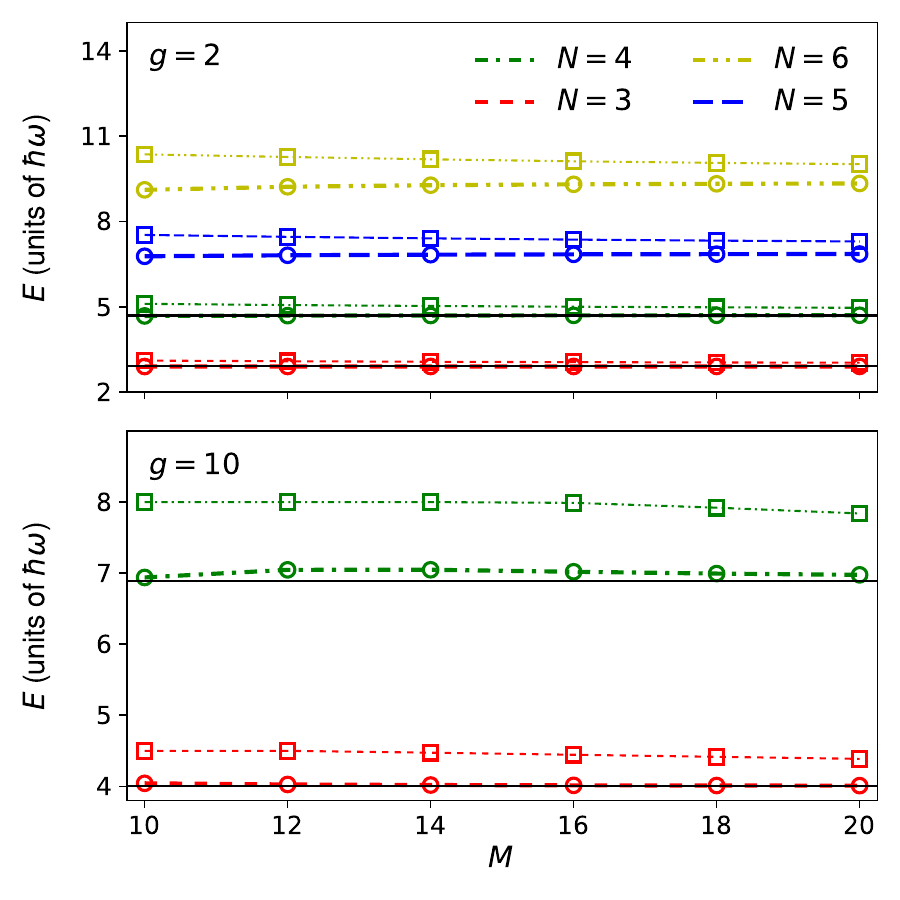}
    \caption{(Color online) Ground-state energy as a function of the number of modes $M$
    for $g=2$ (upper panel) and $g=10$ (lower panel), with $g$ in units of $(\omega\hbar^3/m)^{1/2}$. The squares correspond to results from a truncated diagonalization, whereas the circles correspond to corrected results. The horizontal black dashed lines show the exact energies reported in Ref.\cite{laird_su_2017}}
    \label{sec:con;fig:convergence}
\end{figure}
In a standard truncation in the number of modes without any energy restriction, one simply considers all the states that satisfy $n_i\leq (M-1)\,\forall i$.
This results in a Hilbert space of dimension $M^N$, growing extremely quickly with $M$. In contrast, within the energy truncation we consider all the states with non-interacting energy smaller or equal than $E_\text{max}(M)$ [Eq.~(\ref{sec:ap;subsec:trunc;eq:Emax})], that is, the states which satisfy $\sum_i n_i \leq M-1$. With this truncation scheme
we can work with a much smaller dimension of the Hilbert space without affecting too much the quality of the results~\cite{plodzien_numerically_2018,sowinski_efficient_2019}. Furthermore, and as discussed in Sec.~\ref{sec:ap;subsec:trunc}, the energy truncation allows us to correctly connect the many-body basis with the exact two-body solution.

To compare the sizes of the bases obtained with the two truncation schemes, in Table~\ref{tab:Hilbert_dimension} we show the number of states in the many-body basis obtained with both schemes for number of particles and modes used throughout this article. We observe that, while the basis with the energy truncation grows significantly with both $N$ and $M$, it grows much slower than with the standard truncation.
\section{Convergence of the method}
\label{app:con}
To further illustrate how the calculations depend on the number of modes $M$, in Fig.~\ref{sec:con;fig:convergence} we show how the ground-state energy behaves as a function of $M$ for weak and strong repulsion. We compare the results obtained from the truncated
diagonalization (squares) with the ones including the correction (circles).

The corrected results depend weakly on $M$, showing that our correction produces similar results for a different number of modes, as expected. In particular, we see that for three and four particles the corrected results show only small deviations with respect to the exact solutions (black dashed lines). In contrast, the uncorrected results show important deviations from the exact and corrected results, especially for strong repulsion.

It is also worth noting that the uncorrected energies decrease with $M$ as expected from the variational principle. In contrast, the corrected energies can increase for some choices of $M$, as clearly seen with four particles in the lower panel. This means that the corrected results do not provide an upper bound for the energies. Despite this, the correction provides much more accurate results than the original truncation, making it a preferable choice in direct diagonalization calculations.

\bibliography{On_direct_diagonalization.bib}

\end{document}